\documentclass{ws-procs9x6}

\begin{document}

\title{SPIN LIGHT IN NEUTRINO TRANSITION BETWEEN DIFFERENT MASS STATES}

\author{ A. GRIGORIEV \footnote{ax.grigoriev@mail.ru}}

\address{Skobeltsyn Institute of Nuclear Physics, Moscow State University,\\
119991 Moscow, Russia}

\author{ A. LOKHOV \footnote{lokhov.alex@gmail.com}}

\address{Department of Quantum Statistics and Quantum Field Theory, Moscow State University,\\
119991 Moscow, Russia}

\author{ A. STUDENIKIN \footnote{studenik@srd.sinp.msu.ru}}

\address{Department of Theoretical Physics, Moscow State University,\\
119991 Moscow, Russia}

\author{ A. TERNOV \footnote{a\_ternov@mail.ru}}

\address{Department of Theoretical Physics, Moscow Institute for Physics and Technology, \\
141700 Dolgoprudny, Russia}

\begin{abstract}
The spin light of neutrino is considered in the process of a neutrino
radiative transition between two different mass states in presence
of medium. By this study we investigate the influence
of background matter on the initial and final neutrino states in
the process of massive Dirac neutrino decay due to the non-zero
transition magnetic moment. We derive corresponding corrections
to the total width of the process over the matter density in most
important for applications cases.
\end{abstract}

\bodymatter

\vspace{-0.4cm}

\section{Introduction}\label{aba:sec1}
Electromagnetic properties of neutrino are among the key items of
modern particle physics (see \cite{GiuStu09} for a recent review).
It seems quite natural that a massive neutrino would have nonzero
diagonal or transition magnetic moment. If a neutrino has
non-trivial electromagnetic properties, then neutrino coupling to
photons is possible and several important, for instance for
astrophysics, processes can exist ~\cite{Raf_book96_RafPR99} .
Recently we have proposed a new mechanism of neutrino radiation of
photons that is realized when a relativistic neutrino with nonzero
magnetic moment moves in dense matter. This mechanism was termed
the ``spin light of neutrino" ($SL\nu$)~\cite{LobStuPLB03} . The
quantum theory of this phenomenon was developed
in~\cite{StuTerPLB05} (see also ~\cite{Lob05} and
~\cite{StuJPA_06_08} ).

In this paper we extend our studies of $SL\nu$ ~\cite{StuTerPLB05}
and consider the $SL\nu$ in a more general case when the photon is
emitted in the neutrino radiative decay. The $SL\nu$ considered in
~\cite{StuTerPLB05} was investigated under condition of equal
masses for the initial and final neutrino states. Here below we
examine the case when the neutrino transition between two
different neutrino mass states is realized. Thus, we consider the
$SL\nu$ mode in the neutrino radiative decay in matter originated
due to the neutrino transition magnetic moment.

It should be noted that the neutrino radiative decay was
considered before by several authors ~\cite{Smi78} . It was shown
that the process characteristics are substantially changed if the
presence of a medium is taken into account. In these calculations,
the influence of the background matter was considered only in the
electromagnetic vertex. Here we are going to discuss the impact of
the medium also onto the state of neutrino itself. At the same
time we will be interested in the another aspect of the problem
and consider it from the point of view of light emission. Under
the condition of equal initial and final particle masses the
process becomes equivalent to the $SL\nu$ in matter . With
different masses for the initial and final neutrino states, the
spin light becomes only the constituting channel for the overall
process corresponding to the change of the neutrino helicity. The
mechanism of $SL\nu$ is based on helicity states energy difference
of the particle arising due to the weak interaction with the
background matter. Hence, our study makes sense, obviously, if the
scale of neutrino mass difference is of the order of spin energy
splitting owing to the interaction with matter.

Let us specify now the process under consideration. We are
considering the decay of one neutrino mass state $\nu_1$ into
another mass state $\nu_2$ assuming that $m_1>m_2$, and restrict
ourselves with only these two neutrino species and accordingly
with two flavour neutrinos. Having in mind that conditions for the
most appropriate application of the process under study can be
found in the vicinity of neutron stars we will take for the
background a neutron-rich matter. In this case a process with
participation of antineutrinos is more appropriate and thus will
study here. However for the convenience in what follows we will
still refer to the particles as to neutrinos. Since the
interactions of flavour neutrinos with neutron star matter are the
same and governed by the neutron density we will take equal
interactions for the initial and final massive neutrinos with the
matter.

\vspace{-0.4cm}

\section{Modified Dirac equation}

The system ``neutrino $\Leftrightarrow$ dense matter" depicted
above can be circumscribed mathematically in different ways. Here
we use the powerful method of exact solutions, discussed in a
series of our papers ~\cite{StuJPA_06_08} . This method is based
on solutions ~ \cite{StuTerPLB05} of the modified Dirac equation
for neutrino in the background medium
\begin{equation}
    \{i\gamma_{\mu}\partial^{\mu}-\frac{1}{2}\gamma_{\mu}(1+\gamma^{5})f^{\mu}-m\}\Psi(x)=0,
\label{eq:dirac}
\end{equation}
where in the case of unpolarized and nonmoving matter
$f^{\mu}=G_{f}/\sqrt{2} \ (n,\textbf{0})$ with $n$ being matter
number density. At this, the energy spectrum of neutrino is given
by
\begin{equation}
    E_\varepsilon=\varepsilon\sqrt{(p-s\alpha
    m_{\nu})^{2}+m_{\nu}^{2}}+\alpha m_{\nu}, \ \ \alpha =
\frac{1}{2\sqrt{2}}G_F\frac{n}{m_{\nu}} \label{eq:spektr}
\end{equation}
where $\varepsilon=\pm1 $ defines the positive and negative-energy
branches  of the solutions, $s$ is the helicity of neutrino, $p$
is the neutrino momentum. The exact form of the solutions
$\Psi_{\varepsilon,p,s}(\textbf{r},t)$ can be found in
~\cite{StuTerPLB05} and ~\cite{StuJPA_06_08} .

\vspace{-0.4cm}

\section{Spin light mode of massive neutrino decay}

The S-matrix element for the decay has the standard form that of
the magnetic moment radiation process:
\begin{equation}
    S_{fi}=-(2\pi)^4\mu\sqrt{\frac{\pi}{2wL^3}}\delta(E_2-E_1+w)
    \delta^{3}({\bf p}_2-{\bf p}_1+{\bf k})
    \overline{u}_{f}({{\bf e}},{\bf \Gamma}_{fi})u_i.
\label{eq:amplitude}
\end{equation}
Here ${\bf \Gamma}=i\omega\big\{\big[{\bf \Sigma} \times {\bm
\varkappa}\big]+i\gamma^{5}{\bf \Sigma}\big\}$, $u_{i,f}$ are the
spinors for the initial and final neutrino states, ${\bf e}$ is
the photon polarization vector, $\mu$ is the transitional magnetic
moment~\cite{GiuStu09} and $L$ is the normalization length.

In the process, we have the following conservation laws:
\begin{equation}
    E_1=E_2+\omega; \ \   \bf{p_1}=\bf{p_2}+\bf{k}.
\label{eq:conservation}
\end{equation}
It is useful to carry out our computations through non-dimensional
terms. For that purpose we introduce the following notations:
$\gamma=\frac{m_1}{p_1};\kappa=\frac{\alpha
m_1}{p_1}=\frac{\tilde{n}}{p_1};\delta=\frac{\triangle
m^2}{p_{1}^{2}}=\frac{m_{1}^{2}-m_{2}^{2}}{p_{1}^{2}}$. To single
out the the spin light part of the radiative decay process we
should choose different helicities for the initial and final
neutrinos. Keeping the analogy with the usual process of $SL\nu$
we take the helicity quantum numbers as $s_1=-1$, $s_2=1$. Then
the solution of the kinematic system (\ref{eq:conservation}) can
be written in the form
\begin{equation}
    w=\frac{-(KD+x\kappa^2)+\sqrt{(KD+x\kappa^2)^2-(K^2-\kappa^2)(D^2-\kappa^2)}}{(K^2-\kappa^2)}
\label{eq:freq}
\end{equation}
with the notations $D=s_{1}\kappa-\delta$;
$\tilde{n}=\frac{1}{2\sqrt{2}}G_F n$,
$K=\sqrt{(1-s_{1}\cdot\kappa)^2+\gamma^2}-x$, here $x$ stands for
$\cos\theta$, $\theta$ is the angle between ${\bf p}_1$ and $\bf
k$.

Performing all the calculations we obtain angle distribution of
the probability for the investigated process:
\begin{equation}
    \frac{d\Gamma}{dx}=\mu^2 p_1^3\frac{(K-w+x)(wK-\kappa-\delta)w^3S'}
    {\sqrt{(KD-w+x)^2-(K^2-\kappa^2)(D^2-\kappa^2)}},
\label{eq:density}
\end{equation}
where $S'= (1+\beta_1 \beta_2)(1-\frac{w-x-w \cdot x + w \cdot
x^2}{\sqrt{1+w^2-2w \cdot x}}x)-(\beta_1 +
\beta_2)(x-\frac{w-x}{\sqrt{1+w^2-2w \cdot x}})$ and
$\beta_1=\frac{1+\kappa}{\sqrt{(1+\kappa)^2+\gamma^2}}$,
$\beta_2=\frac{\sqrt{1+w^2-2w \cdot x}-\kappa}{K-w+x}$.

The total probability can be computed from the equation
(\ref{eq:density}) by taking the integration over the angle
$\theta$ range. However, manual calculations are not quite simple
to carry through them. Even though the integral can be calculated
exactly, the final expression is enormously complex and its
explicit form is optional to be given here.

\vspace{-0.4cm}

\section{Results and discussion}
It is worth to investigate the asymptotical behavior of the
probability $\Gamma$ in three most significant relativistic
limiting cases keeping only the first infinitesimal order of small
parameters. On this way we have,
   \begin{equation}
    \Gamma=4\mu^2\tilde{n}^3(1+\frac{3}{2}\frac{m_1^2-m_2^2}{\tilde{n}p_1}+\frac{p_1}{\tilde{n}}),  \
    ({\text {ultrahigh density:}}\   1 \ll \frac{p_1}{m_1} \ll \frac{\tilde{n}}{p_1});
\label{ultrahighdensity}
\end{equation}
\begin{equation}
    \Gamma=4\mu^2\tilde{n}^2p_1(1+\frac{\tilde{n}}{p_1}+\frac{m_1^2-m_2^2}
    {\tilde{n}p_1}+\frac{3}{2}\frac{m_1^2-m_2^2}{p_1^2}), ({\text {high density:}}\
    \frac{m_1^2}{p_1^2}\ll \frac{\tilde{n}}{p_1} \ll 1);
\label{highdensity}
\end{equation}
\begin{equation}
    \Gamma\approx\mu^2 \frac{m_1^6}{p_1^3}, ({\text {quasi-vacuum case:}}
    \frac{\tilde{n}}{p_1} \ll \frac{m_1}{p_1}\ll 1, m_1 \gg m_2).
\label{lowdensity}
\end{equation}

The obtained results (\ref{ultrahighdensity}) and
(\ref{highdensity}) exhibit the power of the method of exact
solutions since they establish clear connection between the case
of massive weak-interacting particles when the masses of the
initial and final particles differ with the previously
investigated equal mass case. Indeed, it is easy to verify that
these results transforms exactly into the results of $SL\nu$
calculation~\cite{LobStuPLB03}. The asymptotic estimation
(\ref{lowdensity}) can not be reduced to the $SL\nu$ case and thus
it is a new result, which is characteristic feature for the decay
process under study. The above-mentioned asymptotical cases
(\ref{ultrahighdensity}), (\ref{highdensity}) and
(\ref{lowdensity}) where calculated with the assumption that the
initial neutrino is relativistic $(\gamma=\frac{m_1}{p_1}\ll 1)$.
In particular the relativistic character of the initial neutrino
propagation influences strongly on the emitted $SL\nu$ photon
energy because of increase of the part of neutrino energy in it.

It is also interesting to investigate the spin light mode in the
radiative decay of slowly moving massive neutrino (or even
stationary initial neutrino). This process has been calculated
several times \cite{Smi78} . We consider the vacuum case to find
the interrelation of our results obtained using the method of
exact solutions with the results of previous works. So, taking
into account $\gamma=\frac{m_1}{p}\ll 1,
\kappa=\frac{\tilde{n}}{p_1}, \delta\equiv\frac{\gamma^2}{2}$ for
the probability of the process we finally get:
\begin{equation}
    \Gamma\approx \frac{7}{24}\mu^2 {m_1^3} \sim m_1^5.
\label{vacuum}
\end{equation}
We obtain here the same dependency of the probability from the
mass of the decaying neutrino as in the classical papers on the
radiative neutrino decay. By this means we justify usage of the
modified Dirac equation exact solutions method.

{\it Acknowledgements.} One of the authors (A.S.)  is thankful to
Kim Milton for the invitation to attend the 9th Conference on
Quantum Field Theory Under the Influence of External Conditions
(Oklahoma, USA) and for the kind hospitality provided in Norman.

\end{document}